\documentstyle[12pt,aasms4]{article}
\begin{document}

\newcommand{\be}{\begin{equation}}
\newcommand{\ee}{\end{equation}}
\newcommand{\noi}{\noindent}
\newcommand{\mdot}{\mbox{$\dot{M}$}}
\newcommand{\msun}{\mbox{$M_{\odot}$}}
\newcommand{\yr}{\mbox{${\rm yr}^{-1}$}}
\newcommand{\kms}{\mbox{${\rm km} \;{\rm s}^{-1}$}}
\newcommand{\ve}{\mbox{${v}_{e}$}}
\newcommand{\Mch}{\mbox{${M}_{ch}$}}
\newcommand{\Efif}{\mbox{${E}_{51}$}}
\newcommand{\Me}{\mbox{${M}_{e}$}}
\newcommand{\Tprime}{\mbox{${T}^{\prime}$}}
\newcommand{\rpr}{\mbox{${R}^{\prime}$}}
\newcommand{\no}{\mbox{${n}_{0}$}}
\newcommand{\Rprime}{\mbox{${R}^{\prime}$}}
\newcommand{\dam}{\mbox{${\rho}_{am}$}}

\title{The Interaction of Type Ia Supernovae with their Surroundings:
The Exponential profile in 2D}

\author{ Vikram V. Dwarkadas}
\affil{RCfTA, School of Physics, A28 Univ of Sydney,  NSW 2006, Australia}

\begin{abstract}

	The evolution of Type Ia supernovae in the surrounding medium
	is studied using 2-dimensional numerical hydrodynamic
	simulations. The ejecta are assumed to be described by an
	exponential density profile, following the work of Dwarkadas
	\& Chevalier (1998). The case of a circumstellar region formed
	by mass loss from the progenitor or a companion star is also
	considered. The decelerating contact discontinuity is found to
	be Rayleigh-Taylor (R-T) unstable, as expected, and the nature
	of the instability is studied in detail for 2 cases: 1) a
	constant density ambient medium, and 2) a circumstellar medium
	whose density goes as $r^{-2}$.  The nature of the instability
	is found to be different in both cases. In the case of a
	circumstellar medium the instability is much better resolved,
	and a fractal-like structure is seen. In the case of a
	constant density medium the extent of growth is less, and the
	R-T fingers are found to be limited by the presence of
	Kelvin-Helmholtz mushroom caps at the tips of the fingers. The
	unstable region is far enough away from the reverse shock that
	the latter is not affected by the mixing taking place in the
	interaction region. In contrast the reverse shock in the case
	of a circumstellar medium is found to be rippled due to the
	formation of instabilities. In neither case is the outer shock
	front affected. These results are consistent with similar
	studies of power-law ejecta profiles conducted by Chevalier,
	Blondin \& Emmering (1992). Our results are then applied to
	Tycho's supernova remnant. We conclude that it is unlikely
	that the instabilities seen in recent radio images of the
	remnant are similar to those studied herein, although we do
	not discount the possibility of initial conditions different
	from those studied herein leading to a much larger growth then
	we see in our simulations. We suggest that such instabilities
	may be better observed at X-ray wavelengths which probe the
	high-density shocked ejecta region.
\end{abstract} 


\section{INTRODUCTION}

The evolution of supernova remnants (SNRs) has been classically
assumed to go through 4 stages (see for example Woltjer 1972). In the
initial free-expansion or ejecta-dominated stage, the expansion of the
remnant is determined by the properties of the supernova ejecta and
the density of the surrounding medium. The mass of the ejected
material, the kinetic energy of the explosion, the density profile of
the ejected material and the density distribution in the ambient
medium are all required to understand the expansion of the remnant in
this stage. This phase is particularly important because many of the
so-called ``historical'' supernovae (SNe) known in our galaxy may
either be in this stage, or be making a transition from it to the
Sedov or adiabatic phase, where the mass swept-up by the SN shock wave
substantially exceeds the ejecta mass.

Determination of the density profile of the ejected material is not a
simple task, and little observational evidence is available. The
density depends on the structure of the progenitor star and the shock
acceleration of the gas during the explosion (Chevalier \& Fransson
1994). The best available information is for SN 1987A, where the
density seems to be well constrained by models of the explosion
(eg. Arnett [1988] and Shigeyama \& Nomoto [1990]). In this case the
density in the outer parts of the ejecta in the free-expansion phase
varies approximately as a power-law with velocity. Similar models have
been used to describe Type II SNe in general, and self-similar
solutions for power-law ejecta interacting with a power law or
constant density surrounding medium have been derived by Chevalier
(1982a) and Nadozhin (1985). Recently, Truelove \& McKee (1999) have
presented more general approximate solutions for the evolution of
non-radiative SNRs in the ejecta-dominated stage. The special case of
interaction of a uniform density ejecta with the surrounding medium
will also be addressed in Drury and Dwarkadas (2000).

Power-law models (with an exponent of 7) have also been used to
describe the ejecta density profile in Type Ia SNe (Chevalier 1982a,
1982b; Fabian, Brinkmann \& Stewart [1983], Band \& Liang [1988]),
which are believed to result from the degenerate ignition of a C/O
white dwarf in a binary system. However Dwarkadas \& Chevalier (1998,
hereafter DC98) showed that an exponential ejecta profile for Type Ia
SN explosions gives the best fit to the density profile. Furthermore,
they compared the interaction of Type Ia SNe described by power-law,
exponential and constant density ejecta profiles with the surrounding
medium, in the simplest case of spherical symmetry.

This paper expands on the work of DC98, presenting high-resolution
two-dimensional simulations of ambient medium interaction in Type Ia's
using an exponential density profile to describe the ejecta. We
compare the properties of the two-dimensional flow to that in
one-dimension. The 2D grid allows us to investigate the
multi-dimensional aspects of the flow, especially the formation of
hydrodynamic instabilities in the free-expansion phase. In particular
we study the evolution and growth of the Rayleigh-Taylor (R-T)
instability that arises at the decelerating contact discontinuity.

The role of the R-T instability in young supernova remnants has been
receiving great attention in recent years (see Chevalier 1996). The
accumulation of observational evidence of substantial non-radial
motions and mixing occurring in the explosion of SN 1987A led several
groups to perform multi-dimensional simulations of supernova
explosions (Arnett et al.~1989; Benz \& Thielemann 1990; Fryxell et
al.~1991; Herant \& Benz 1991).  Hydrodynamic instabilities in
self-similar driven waves were investigated by Chevalier, Blondin \&
Emmering (1992, hereafter CBE92). The presence of instabilities can
lead to fragmentation and the formation of clumps or blobs in the
interface between the ejecta and the surrounding material. Mixing
among various layers can destroy any early stratification and cause
certain elements to be found at larger radii than would otherwise be
expected. In this paper our aim is to characterise the nature and
growth of these instabilities for the case of SN ejecta described by
an exponential density distribution.  Implications of mixing of the
ejecta and circumstellar (CS) material for the observable supernova
remnants, in particular Tycho's SNR (SN1572), is also addressed. The
growth of these instabilities under various conditions is
investigated, such as a constant density medium (indicative of the
interstellar medium) or a circumstellar medium formed by mass loss
from the progenitor star. We compare and contrast the results in these
two cases.

The rest of this paper proceeds as follows. In \S 2 we first review
briefly the properties of an exponential ejecta density profile. This
is followed by initial results of the 2D simulations in \S 3. In \S 4
we discuss the R-T instabilities in more detail. \S 5 focuses on the
application of our results to Type Ia SNe, especially Tycho's SNR,
while \S 6 contains a brief discussion and summary.

\section{The Exponential Ejecta Profile}

The exponential profile used to describe the SN ejecta can be written
as (DC98):

\begin{equation}
{\rho}_{SN} = A \exp (-v/v_e)\;t^{-3}
\end{equation}

\noi where, by integrating over the total ejecta distribution, taken
to have mass ${ M}_e$ and kinetic energy ${ E}$, we find that

\be
\ve = \left[ {{E} \over {6 \Me}}\right]^{1/2} = 2.44 \times 10^8 {\Efif}^{1/2} \left[{\Me \over \Mch}\right]^{-1/2} {\rm cm} \; {\rm s}^{-1} \;,
\ee

\noi 
where $\Mch \equiv 1.4\; M_{\odot}$ is approximately the Chandrasekhar
mass, and $\Efif$ is the explosion energy in units of 10$^{51}$
ergs. The value of $\Efif$ quoted in the literature lies anywhere
between about 0.3 and 1.8. The parameter $A$ is given by:

\be
A = \frac{6^{3/2}}{8 \pi} {{{\Me}^{5/2}} \over {E^{3/2}}} = 7.67 \times
10^6 \left[ {{\Me} \over \Mch}\right]^{5/2} {\Efif}^{-3/2}\; {\rm g}\; {\rm s}^{3} \; {\rm 
cm}^{-3} \;.
\ee

\noi The interaction of the SN ejecta with the ambient medium gives
rise to a double shocked structure, consisting of a forward shock
expanding into the surrounding medium, and a reverse shock that moves
inward in a Lagrangian sense, separated by a contact discontinuity. We
have carried out 2-dimensional hydrodynamic simulations of the
supernova ejecta colliding with the ambient medium. The assumption of
spherical symmetry with regard to the exponential ejecta density
distribution is certainly an approximation. In reality the formation
of instabilities during the explosion will probably destroy any such
early symmetry, leading to large-scale, possibly turbulent convective
motions (M$\ddot{\rm u}$ller 1998). Nevertheless, this assumption is a
valid starting point for such a study, and the inclusion of clumps and
non-radial flows would be the subject of future, more complicated
studies (see Wang \& Chevalier 2000)

Our lack of knowledge of the progenitors of Type Ia SNe (eg.~Livio
1999) makes it unclear what the density structure is in the immediate
surroundings. The lack of radio and X-ray emission, and of nebular
excitation lines from circumstellar interaction, points to a medium of
constant low density. However if Type Ia's are formed from degenerate
ignition in a white dwarf, then it is clear that the white dwarf must
be in a binary system, and mass transfer must be occurring (Wheeler
1996). A circumstellar region due to mass loss from the secondary star
would then be expected, although its extent may be small. Cumming et
al.~(1996) place limits on the properties of such a region. In this
paper we consider both cases, that of a constant density surrounding
medium and a circumstellar medium. If the mass loss rate and velocity
of the wind from the companion star are constant, the density of the
circumstellar medium (CSM) decreases inversely as the square of the
radius. The supernova remnant will remain in such a medium only for a
few decades at most. Nevertheless, as shown by DC98, the passage of
the shock wave through such a CS region can leave an impression on the
density structure of the interaction region even after the shock wave
has exited the CS region. Thus it is both observationally relevant as
well as academically interesting to study the evolution of the shock
within a circumstellar medium. Furthermore, while an exponential
profile has at present been postulated only in relation to Type Ia
SNe, it must be emphasised that it is really not clear what the
structure of the ejecta density profile is for any supernova other
than possibly SN 1987A, despite recent progress in this direction (see
for instance Matzner \& McKee [1999]). A detailed study of the
exponential ejecta density expanding into various ambient media is
thus warranted.

\section{2D Simulations}

The simulations were carried out using the VH-1 code, a 3-Dimensional
finite-difference hydrodynamic code based on the Piecewise Parabolic
Method (Colella and Woodward 1984), written by the Virginia Hydro
group. The code, obtained from Dr.~John Blondin at NCSU, solves the
equations of conservation of mass, momentum and energy in Lagrangian
co-ordinates, and then remaps the variables back to the original
Eulerian grid. It has been extensively tested on various astrophysical
hydrodynamical problems. The version employed here uses an expanding
grid that tracks the outer shock front and expands along with it,
making it ideally suited to problems where the dimensions change by
many orders of magnitude over the course of the run.

The evolution of the outer shock front in 2D is similar to that in 1D,
as described by DC98. One important quantity is the evolution of the
expansion parameter with time. If the radius is proportional to time
raised to some power (R $\propto$ t$^{\delta}$), then $\delta$ is the
expansion parameter. This is a dimensionless quantity that can be used
to discriminate between the various phases of remnant evolution. In
Figure 1 we show the evolution of the expansion parameter with time
for a Type Ia SN evolving in a constant density medium. The time axis
is normalised to the quantity $\Tprime \approx 248 \, {\Efif}^{-0.5}
\left( {\Me \over {\Mch}} \right)^{5/6} {\no}^{-1/3}~ {\rm yr}$ using
equation (4) in DC98. The number density $\no$ is related to the
ambient medium density $\rho$ by $\no$ = $\rho$/2.34 $\times 10^{-24}$
cm$^{-3}$, appropriate for a medium with a H/He ratio of 10:1. In
Figure 2 we show the same plot for a CS medium whose density decreases
as $r^{-2}$. In this case the time normalisation is taken to be
${\Tprime}_2 = E^{-0.5} {M_e}^{1.5} D^{-1}$, where $D = \rho
r^2$. These plots show how the expansion parameter varies over the
evolution of the remnant in the free expansion phase, until it reaches
the Sedov value, which is 0.4 in the case of the constant density
medium and 0.67 in the case of a circumstellar medium. The continuous
evolution of the expansion parameter is consistent with the
higher-resolution 1D calculations of DC98.

As the outer shock expands into the circumstellar medium, it begins to
sweep up the ambient material. The Sedov value of the expansion
parameter in either case is reached only once the swept-up mass
significantly exceeds the ejecta mass, by a factor of 30 or
more. However, once the mass of swept-up material becomes comparable
to the ejecta mass, the pressure of the surrounding medium causes the
SN ejecta to decelerate. As the mass of shocked ambient material
continues to increase, the reverse shock begins to move inwards
towards the center, and the contact discontinuity between the two
shocks begins to decelerate. The high density shocked SN ejecta behind
the contact discontinuity are decelerated by the swept-up material
ahead of it. The pressure and density gradients lie in opposite
directions, and the system becomes unstable to Rayleigh-Taylor (R-T)
instabilities.  Small-amplitude perturbations, with wavelength on the
size of a few grid zones, then occur. The size of these perturbations
appears to grow linearly, and soon the familiar R-T fingers spread out
into the shocked surrounding medium. The tips of these fingers
terminate in bulbous-shaped mushroom-like heads, a result of the
Kelvin-Helmholtz instability as the shocked gas surrounding the
fingers tries to slide past them. This evolution is depicted in
Figures 3 and 4, which are discussed in detail below.

\section{The Rayleigh-Taylor instability}

The Rayleigh-Taylor instability arises at the interface between a
low-density fluid supporting a high-density fluid (e.g. Chandrasekhar
1961). In our case the analogous situation arises at the contact
discontinuity separating the shocked SN ejecta from the shocked
ambient medium, as described above. Small ripples arise initially
along the entire length of the contact discontinuity. No explicit
perturbation is required to induce the instability, which can be
triggered purely by the numerical noise in the system. We have also
tested the case when explicit perturbations of various amplitudes are
introduced (\S 4.2). The ripples in our higher resolution simulations
appear to start off close to the symmetry axis and then move towards
the equator, but this is not easily confirmed, since by the time the
ripples acquire an appreciable amplitude they are spread over the
entire contact discontinuity. It is however quite likely that
approximations at the boundary lead to minor fluctuations that trigger
the instability.

Figure 3 shows grey-scale images of the evolution of the instability
with time, for a SN evolving in a constant density medium, which we
refer to henceforth as Case 1. Figure 4 shows the same for a Type Ia
SN evolving in a circumstellar medium, referred to as Case 2. The
times listed for each plot are normalised in the same manner as for
the expansion parameters (\S 3). The radius for Case 1 is normalised
to the value $\Rprime = \left( {{\Me } \over {\frac{4}{3} \pi \rho}}
\right)^{1/3} \approx 2.19 \left( {\Me \over {\Mch}} \right)^{1/3}
{\no}^{-1/3}~ {\rm pc} $.  For Case 2 the radius normalisation is
$\Rprime={M_e}/D$, where $D = \rho r^2$.  The growth of the finger
size (the penetration of the denser layer into the less dense layer)
with time for Case 1 is shown in Figure 5. The plot for Case 2 looks
quite similar when normalized to the appropriate units and is not
shown.

Theory predicts that the size of the R-T fingers should increase as
the square of the time. However a simple quadratic fit was found to be
inadequate to model the growth, while a second order least-squares
polynomial fit was found to provide a reasonable solution. The
deviations are much larger earlier on, which is partly due to our
inability to determine the amplitude of the perturbation in the very
early stages. The fact that a linear as well as a quadratic component
is needed is consistent with the fact that the flow is not
self-similar (eg. Dalziel et al.~1999).

A standard linear analysis of the R-T instability shows that the
growth rate is proportional to the square-root of the wavenumber
(eg.~Chandrasekhar 1961; Sharp 1984). According to this the smallest
wavelengths should be the most unstable. In our simulations we find
that the initial ripples are on the scale of a few (5-10) grid
zones. Smaller wavelengths are damped by any inherent viscosity in the
numerical method. The ripples grow into spike-like protrusions,
forming the characteristic Rayleigh-Taylor `fingers'. As the spiky
protrusions expand outwards into the surrounding medium, the shocked
ambient material tries to slide past them. This gives rise to shear
flow at the edges, leading to the growth of Kelvin-Helmholtz (K-H)
instabilities, visible as the bulbous, mushroom-shaped heads on the
R-T fingers. The flow of the shocked ambient medium past the fingers
also tends to limit their growth, more so in the constant density
medium case. Over time the fingers appear to merge into one another
and the fastest growing wavelength increases with time, or
correspondingly the number of individual R-T fingers seen decreases
with time. The scale of the fastest-growing wavelength is set by the
system, independent of the initial perturbation that triggered the
instability. The result is that as the reverse shock approaches the
center, the unstable wavelength, or alternatively the number of R-T
fingers seen, is the same no matter what the amplitude of the initial
perturbation, or even if no initial perturbation was applied to the
system.

All our simulations show a jet-like structure along the symmetry
axis. This structure is even more prevalent in those simulations where
an initial perturbation was introduced. When the jet breaks through
the outer shock it tends to ``fan out'', thus giving rise to a large
mushroom-shaped cloud. These structures, prevalent in many other
numerical simulations embodying the same symmetry, are most likely
numerical artifacts and are not real. They are formed at the symmetry
axis, where the geometry results in grid zones whose volume is very
small. Any fluid component perpendicular to the axis of symmetry
(i.e. flowing into the axis) has nowhere to go and is deflected along
the axis, thus resulting in these spurious structures.

We elaborate later on the differences between the instabilities seen
in the constant density and the 1/$r^{2}$ surrounding medium, after
first discussing the similarities between the two cases. A point to
note in all our simulations is that, irrespective of the initial
amplitude of the perturbation or of the nature of the surrounding
medium, the unstable region never occupies more than about 50\% of the
interaction region. The size of the R-T projections grows faster than
the size of the interaction region, so that the width of the mixing
region tends to increase with time. However in all cases the reverse
shock reaches the center, and the instabilities ceased to grow, when
the mixing width was less than or about half the size of the
interaction region. One implication of this is that the instabilities
never penetrate the outer shock, and in fact have no direct effect on
it. The appearance of the outer shock is not changed by the
considerable amount of mixing occurring in the interaction region
close to the contact discontinuity. The same is not true of the
reverse shock in the circumstellar case, as described later.

While the above results are consistent with those found by CBE92 in
their study of the Rayleigh-Taylor instability for power-law ejecta, a
major difference exists between the two cases. The latter describe the
quasi-steady evolution of the instability in the self-similar driven
case, and in fact show (their Figure 7) how snapshots of the evolution
taken at times two orders of magnitude apart hardly reveal any
difference. The same is not true in the exponential case. Introduction
of the velocity scale length in the exponential profile means that the
solution is no longer self-similar, even though it is scalable. The
appearance of the unstable region tends to change, albeit slowly with
time, and no two snapshots which are two decades apart in time
resemble each other closely as in the models of CBE92.

A close look at Figures 3 and 4 shows that there is considerable
difference between the growth of the instabilities in the two
cases. In the constant density medium case the growth of the fingers
appears stunted. The formation of the mushroom caps at the head of a
finger tends to limit the growth of the projection. The projections
grow to a much larger extent in the CS case, and occupy a greater
fraction of the interaction region. We attribute these differences in
large part to the difference in Atwood number between the two cases
(eg.~Sharp 1984). The Atwood number, or equivalently the density
contrast between the shocked ambient and shocked ejecta material,
governs the growth of the instabilities for small-amplitude
perturbations of wavelength close to the critical wavelength required
for unstable behaviour (Sharp 1984). The Kelvin-Helmholtz instability
caused by the shocked fluid flowing past the spikes is more pronounced
at lower values of the Atwood number, which is the case when the shock
expands in a constant density medium. This can cause the caps to
mushroom out, increasing the effect of the drag on the spike and
slowing its growth. This difference in the two cases is similar to
that seen by CBE92 in their study of R-T instabilities for
self-similar driven waves.

The two cases are further differentiated by the effect of the
instabilities on the reverse (inner) shock. In the constant density
medium case, the contact discontinuity is far enough away from the
reverse shock that the instabilities do not affect the shape of the
shock, which stays smoothly spherical. However in the CS case the
reverse shock is much closer to the contact discontinuity, and the
pressure variations are higher, causing the reverse shock itself to
get wrinkled. The corrugation of the inner shock (Figure 4, middle
panels) occurs on a scale similar to the most unstable wavelength, and
seems to get more pronounced as the reverse shock heads towards the
origin. Again, a similar effect was noticed by CBE92.

Note also that the time taken for the reverse shock to reach the
centre in Figure 3 is about two-thirds of that found in the work of
DC98. This is to be expected. The instabilities cause a ``leakage'' of
pressure from the shocked ambient medium to the shocked ejecta. Thus
the pressure driving the reverse shock is higher than in the
spherically symmetric work of DC98, and the reverse shock consequently
reaches the origin earlier.

The most likely scenario for supernova shock evolution is that it
traverses a region of circumstellar medium followed by a constant
density interstellar medium. As the shock moves from the CS to the
constant density medium, the nature of the instabilities will
change. The details will depend strongly on the extent of the CS
region. If the size of the initial CS region is small, such that the
instabilities do not have time to grow, then the effect of the CS
region on the later evolution of the remnant is small. However we do
find that in most cases the reverse shock remains corrugated even
after the remnant is expanding primarily in a constant density
medium. Thus, in addition to the spike in the density profile at the
contact discontinuity, and the corresponding temperature drop
mentioned in DC98, the shape of the reverse shock itself is affected
by passage through an initial CS region. The presence of a CS shell at
the interface between the two media was not taken into account and
would further affect the evolution of the instabilities.

\subsection{Effect of Grid Resolution}

The effects of grid resolution on the instabilities merits
consideration. Increasing the resolution leads to the appearance of
structure on smaller scales, as would be expected. In Case 2 the
instabilities appear to clump together with a number of sub-fingers
arising from each finger, instead of appearing as individual
spikes. The turbulent mixing behaviour, with vortices being formed
close to the projecting fingers, causes individual fingers to be bent
at an angle, producing structures resembling question marks. The stems
show the usual Kelvin-Helmholtz mushroom caps due to shear flows
around the edges. The number of R-T clumps that exist when the reverse
shock reaches the center remains the same in all simulations, although
the substructure in each clump continually increases. The appearance
somewhat resembles a tree-structure, with higher resolution revealing
more branches in the tree and more stems arising from the branches
(see top panels in Figure 6). The pattern is fractal in nature,
although the code is unable to resolve individual stems very well. The
mixing width (the width of the region occupied by the instabilities)
shows a definite increase of a few percent with resolution in the
earlier stages of the instability. This may be partly due to the fact
that as the shock expands outwards, individual structures are more
easily discernible in a higher resolution simulation as the increase
in the size of a grid zone is not as large as in a lower-resolution
simulation. As the instabilities evolve the mixing width occupied by
the unstable region becomes approximately constant.

In Case 1 a similar overall effect is visible. Simulations with a
higher grid resolution lead to more R-T fingers being visible. As the
simulation evolves the amount of substructure also increases, as would
be expected. Also visible is an increase in the size of the K-H caps
at the heads of the stems in the higher resolution cases. As before
however, the simulations appear quite similar once the reverse shock
nears the inner boundary, and the number of R-T fingers is more or
less the same. No fractal structure is seen, in contrast to case 2,
although the fingers are much better defined with an increase in the
number of grid zones, as would be expected.  Another effect is that
the instabilities appear to start earlier as the spatial resolution
increases. An increase in grid resolution by 50\% leads to a decrease
in the time when the ripples are first seen by about 20\%. Also
important is the fact that, as the growth rate is limited, a minimum
resolution is required to capture the growth of the
instabilities. This is consistent with the results of Fryxell et
al.~(1991) for SN 1987A. We have found that even a 300 $\times$ 300
run does not generate the instabilities till the reverse shock is
close in to the centre of the explosion, and that a grid of at least
500 azimuthal zones is necessary to capture the instabilities in
detail.

\subsection{Effect of initial perturbation}

We have investigated the effect of adding an initial perturbation of
various amplitudes. A sinusoidal perturbation of wavelength equal to
the angular dimensions of the grid was added. Figure 6 shows snapshots
at approximately the same epoch in time from various runs for Case
2. Two cases with no perturbation added but different resolutions are
shown in the first row. The remaining frames show cases where a
perturbation of the given amplitude was introduced, all with the
resolution set to 500 by 500 zones. In general we find that for Case 2
the overall effect of small-amplitude perturbations ($\la 10 \%$) is
tiny, and the instabilities develop in almost the same way as they do
in the case of zero initial perturbation. Initially the differences
appear large, with the no-perturbation simulations showing a more
homogeneous growth of projections whereas simulations with initial
perturbations show fewer and more well-defined fingers. However the
differences decrease as the reverse shock moves inwards, and in most
cases the appearance of the unstable region looks quite similar by the
time the reverse shock nears the inner edge of the grid. The effects
of perturbations of amplitudes $\ga 15\%$ are more pronounced. There
is an increase in the amount of small-scale structure to be seen, and
the larger perturbation simulation presents a more clumpy
appearance. A small increase in the mixing width accompanies the
increase in the amplitude of the perturbation. It is evident from
Figure 6 that the effects of higher resolution and introduction of a
perturbation are not independent, in the sense that increasing the
resolution has a similar effect to adding a perturbation. The higher
resolution case with no perturbation is quite similar to the runs with
an initial perturbation.

For Case I the effect of adding an initial perturbation is more
pronounced, as shown in Figure 7. The inhomogeneous appearance of very
small wavelength projections is replaced by individually discernible
fingers. The projections seem to stretch out further, but their number
has decreased. Thus, although there is a definite increase in the
mixing width, the volume occupied by the instabilities remains about
the same. Again the differences decrease as the simulation proceeds,
although even when the inner shock is close to the center the
differences are immediately perceptible, unlike in Case 2. Another
distinction is that with the addition of an initial perturbation, the
instabilities form much earlier than in the case of no perturbation,
and a lower grid resolution is adequate to show that the instabilities
are present, although a higher resolution grid is still required to
resolve them. A grid with 300 azimuthal zones is sufficient to at
least demonstrate the growth of the instabilities (Fig 7).

\section{Tycho's SNR (SN 1572)}

Rayleigh-Taylor instabilities have been often cited as the cause of
corrugations and rippled structure in images of SNRs. Velazquez et
al.~(1998) have studied the Rayleigh-Taylor instability in Tycho's
SNR, following up on VLA observations carried out by Reynoso et
al.~(1997). They conclude that wavy structure seen in the NE quadrant
of the VLA image, formed by small, regularly spaced protrusions just
behind the outer shock front, is due to Rayleigh-Taylor
instabilities. It is unclear if the results of our work support such
an interpretation. As emphasised in the previous section, in no case
did the instabilities in our simulations cover more than half the
interaction region, let alone penetrate the outer shock front, as do
some of the protrusions shown in Figure 2 of Velazquez et
al.~(1998). If the radio emission is coming from just behind the outer
shock our simulations imply that it is unlikely that any unstable
structure should be visible close to to the outer rim. Moreover, the
calculations of DC98 and our own 2D calculations show that a tenable
model for Tycho requires that the reverse shock be quite far out, and
the size of the interaction region be small. The instabilities will
then still be in the incipient stages, as is also found by Velazquez
et al.~(1998). However that would mean that they would occupy an even
smaller fraction of the interaction region, and would not likely be as
prominent as they are in the radio images, unless some effect led to
an increase in entropy within the mixing region, leading to a larger
growth of the R-T instabilities than is observed in our
simulations. The latter is not entirely improbable - it is possible
perhaps that thermal instabilities in the surrounding medium,
operating over evolutionary timescales, set up a very clumpy and
non-linear initial condition for the growth of these instabilities,
resulting in much larger growth than that studied herein.  There
exists also the possibility that if the NE portion is running into a
high density H1 cloud as revealed by Reynoso et al.~(1997), some of
the corrugated shock structure may be due to interaction of the
remnant with an uneven cloud boundary.

Based on the results of the simulations carried out herein, we may
argue that emission which arises predominantly from just behind the
outer shock is unlikely to reveal evidence of the R-T instabilities
studied in this paper. However the instabilities might show up better
in maps of emission from close to the interface between the shocked
ambient and shocked ejecta gas. A case in point is again Tycho's
SNR. X-ray line observations by Hwang and Gotthelf (1997) have
provided narrow-band images in various emission lines, such as Si, S
and Fe. These reveal a shell which is generally circular, but many of
the images (their Figure 4e) show a corrugated, wavy outer shell and
the presence of clumps. An isolated knot is also seen in several
images near the southeast bulge. The presence of Si, S and Fe lines
indicates that much of the emission is coming from the reverse-shocked
ejecta material (see also Hwang et al.~1998). The Si and S emission
may in fact be arising from close to the contact discontinuity
(DC98). If so, the rippled outer boundaries may be a result of R-T
instabilities similar to those described in this paper. We have
carried out simulations with parameters that may be reasonable for
Tycho's SNR. Although these simulations have not exhaustively explored
the parameter space (and as such are not intended as a specific model
for the remnant), they do indicate that instabilities would arise
within a couple of hundred years after the explosion, and that,
depending on the initial conditions, they would not yet have
progressed into the non-linear phase at the age of Tycho's SNR.

We could perhaps summarise by saying that for SNe with an exponential
ejecta density profile in the free-expansion phase, the presence of
instabilities near the contact discontinuity is more likely to be
revealed in X-ray than in radio images. This implies a lack of
correlation between the relative brightness of the radio and X-ray
images, at various points around the circumference. Such a lack of
correlation is seen in images of Tycho (Hwang and Gotthelf 1997).

\section{Summary}

We have carried out 2-dimensional simulations of Type Ia SNe
interacting with the ambient medium, expanding on the work of DC98. An
exponential density distribution was assumed for the ejecta. The
simulations reveal that the decelerating contact discontinuity
separating the shocked supernova ejecta from the shocked surrounding
medium is unstable to Rayleigh-Taylor instabilities, as expected.

The nature and characteristics of the instabilities was examined. The
instabilities tend to cover at most 50\% of the interaction region. In
no case is the outer shock affected by the turbulent mixing in the
interior. However, if the surrounding medium is a circumstellar medium
formed by mass loss from the companion star, then the reverse shock is
found to be wrinkled due to the effect of the instabilities.  The
scale of the instabilities as they enter the non-linear regime is
independent of the initial perturbations introduced into the system.

The effect of increasing resolution is to increase the small-scale
structure visible in each R-T finger. A fractal-like pattern is seen
for a surrounding CS medium, with multiple fingers forming an R-T
clump. However the number of such clumps remains the same irrespective
of the initial perturbation. If the ambient medium is of constant
density, then the amount of substructure seen is greater, but no
clear, well-defined fractal pattern is formed.

Our results suggest that corrugated outer shocks in SNRs cannot be
attributed to instabilities of the type studied herein. We do caution
however that this may be a result of the initial conditions assumed
herein, and it is possible that different initial conditions, or a
non-axisymmetric surrounding medium, could result in much larger
growth of the R-T fingers. We feel that the instabilities are more
likely to be visible in emission that arises from the reverse-shocked
ejecta, such as X-ray emission from young supernova remnants. This is
perhaps the case for Tycho's SNR, and some of the clumpy structure
seen in the images reported by Hwang and Gotthelf (1997) could
possibly be a result of R-T instabilities.

The study of instabilities helps us to understand the
multi-dimensional structure that is formed during the evolution of
supernova remnants, and thereby aid in interpretation of the
observational data. Further studies are required that take into
account instabilities that may occur during or just after the
explosion process, and determine how these affect the later
evolution. Also, 3-dimensional studies often tend to show subtle
differences from 2-dimensional ones, and a fuller understanding will
require 3-dimensional simulations.

\acknowledgements {\bf Acknowledgements} I would like to thank
Dr.~Roger Chevalier and Dr.~Lewis Ball for useful discussions and
comments on the manuscript, Dr.~Noella D'Cruz for a critical reading,
and the anonymous referee for useful comments.

\clearpage

{\bf Figure Captions}

\figcaption{Evolution of the expansion parameter with time, for a SNR
evolving in a constant density medium. The X-axis denotes time in
normalised units (see text). The expansion parameter is calculated
after every 20 timesteps rather than being averaged over, leading to
somewhat large fluctuations in this and the next figure. The evolution
proceeds until the reverse shock is close to the center and the
expansion parameter approaches the Sedov-Taylor value of 0.4,
appropriate for a constant density ambient medium.}

\figcaption{Evolution of the expansion parameter with time, for a SNR
evolving in a circumstellar medium. The X-axis denotes time in
normalised units (see text). The evolution proceeds until the reverse
shock is close to the center and the expansion parameter approaches
the Sedov-Taylor value of 0.67, appropriate for an ambient medium
whose density decreases as r$^{-2}$.}

\figcaption{Density snapshots from a 500 X 500 run for a Type Ia SNR
evolving in a constant density medium. No initial perturbation was
applied. The size scale of the grid and the time at which each plot is
depicted are listed on each image, in terms of the scaling variables
described in the text. The shading is such that lighter colour
represent higher densities. The innermost and outermost black regions
do not constitute a part of the grid.}

\figcaption{Density snapshots from a 500 X 500 run for a Type Ia SNR
evolving in a CSM with density decreasing as r$^{-2}$. No initial
perturbation was applied. The size scale of the grid and the time at
which each plot is depicted are listed on each image, in terms of the
scaling variables described in the text. The shading is similar to
that in the previous image.}

\figcaption{A plot of the size of the R-T fingers versus (normalised)
time. The dashed line represents a second-order polynomial fit.}

\figcaption{Density snapshots, at approximately the same epoch, from
various runs for case 2. The top row includes two runs with different
grid resolutions without the addition of an initial perturbation. The
other runs included an initial perturbation of different amplitudes,
as labelled, with a 500 $\times$ 500 zones resolution.}

\figcaption{Density snapshots, at approximately the same epoch, from
three different runs for case 1. The leftmost image is from a run
where no initial perturbations were introduced. The latter two
included an initial perturbation but differ in grid resolution.}

\end{document}